\documentclass[twocolumn,english,superscriptaddress,aps,prl,longbibliography,nobibnotes, groupedaddress]{revtex4-1}
\usepackage{xcolor}
\usepackage{bm}
\usepackage{amstext}
\usepackage{amssymb}
\usepackage{graphicx}
\usepackage{commath}
\usepackage{dsfont}
\usepackage{mathtools}
\usepackage{amsmath,mathrsfs}
\usepackage{physics}
\usepackage{braket}
\usepackage{hyperref}
\usepackage{algpseudocode}
\usepackage{algorithm}
\usepackage{comment}
\usepackage{soul}
\usepackage{verbatim}

\newcommand{\prlsection}[1]{\textit{#1---}\ignorespaces}

\begin{document}
\title{Universal Counterdiabatic Quantum Sensing}
\author{Stewart Morawetz}
\email{morawetz@bu.edu}
\affiliation{Department of Physics, Boston University, Boston, Massachusetts 02215, USA}

\author{Anatoli Polkovnikov}
\affiliation{Department of Physics, Boston University, Boston, Massachusetts 02215, USA}

\begin{abstract}

One of the primary applications of quantum mechanics is leveraging coherence to make incredibly sensitive measurements. One way to push the boundary of this sensitivity even further is to initially prepare a system in an entangled state, though the fragility of these states limits their usefulness. In this work, we propose a new approach by which sensitivity comes \textit{during} the preparation of an entangled state, rather than first preparing the state and then using it for sensing. An adiabatic state preparation process is then accelerated by universal counterdiabatic driving, which is more robust to uncertainty in microscopic parameters than standard shortcuts to adiabaticity. Using a toy Lipkin-Meshkov-Glick model, we show that this approach can also lead to sensitivity beyond the Standard Quantum Limit.

\end{abstract}

\maketitle

\prlsection{Introduction} Quantum sensing \cite{degenQuantumSensing2017} is one of the most important applications of quantum mechanics. Quantum-enhanced measurement precision has been demonstrated experimentally and is actively used in technology. In each case, a quantity to be measured can be encoded in the relative phase between states in coherent superposition, allowing it to be determined with high sensitivity. Enormous technical progress in the realization and control of the well-isolated quantum systems necessary for sensing has been made in recent years ~\cite{georgescuQuantumSimulation2014, wendinQuantumInformationProcessing2017, morgadoQuantumSimulationComputing2021, foss-feigProgressTrappedIonQuantum2025}.

Traditionally, gains in sensitivity have come from making many independent measurements. If a measurement is repeated (or performed in parallel) $N$ times, the uncertainty in the estimate of the quantity to be measured decreases by a factor of $1/\sqrt{N}$, leading to a lower bound on uncertainty known as the Standard Quantum Limit (SQL). If these measurements are instead made on $N$ correlated particles, such as in the case of squeezed light \cite{cavesQuantummechanicalNoiseInterferometer1981} or spins \cite{kitagawaSqueezedSpinStates1993a, winelandSpinSqueezingReduced1992}, then uncertainty in the estimate may instead be decreased by up to a factor of $1/N$, a lower bound known as the Heisenberg limit \cite{giovannettiQuantumEnhancedMeasurementsBeating2004, boixoGeneralizedLimitsSingleParameter2007}.

Most approaches to achieving this enhanced scaling with $N$ require preparing the system in an entangled (correlated) state. There has been considerable effort in this direction beyond squeezed states, for example in attempts to realize GHZ \cite{greenbergerBellsTheoremInequalities1990,bollingerOptimalFrequencyMeasurements1996, kesslerHeisenbergLimitedAtomClocks2014,friisObservationEntangledStates2018a,omranGenerationManipulationSchrodinger2019} or N00N \cite{kokCreationLargephotonnumberPath2002,mitchellSuperresolvingPhaseMeasurements2004,dowlingQuantumOpticalMetrology2008,dengisVortexNOONStates2026} states. Both are quite susceptible to dephasing noise \cite{ huelgaImprovementFrequencyStandards1997a, bohmannEntanglementPhaseProperties2015} and hence in most realistic scenarios the theoretically optimal gain cannot be achieved.

Another direction of research has been quantum critical metrology, \cite{zanardiQuantumCriticalityResource2008, frerotQuantumCriticalMetrology2018, mihailescuCriticalQuantumSensing2026} where the diverging susceptibility to perturbations of a ground state at criticality has been suggested as a highly sensitive measurement probe. These states enable beyond-SQL sensitivity, even after the time required to prepare them adiabatically has been properly accounted for \cite{ramsLimitsCriticalityBasedQuantum2018}. Although the theoretical sensitivity gain is enticing, practical challenges associated with the control of many-body quantum systems near criticality remain.

In this work, we propose a new approach for quantum sensing, whereby the presence of a coupling is inferred via its effect on an adiabatic process, sped up by a shortcut to adiabaticity which does not depend on the parameter to be determined. Using these shortcuts to prepare quantum critical states for metrology has been attempted before \cite{gietkaAdiabaticCriticalQuantum2021}, but the gains have been limited by conventional shortcut approaches requiring knowledge of the to-be-determined parameter beforehand. Our approach differs in that we are not attempting to prepare states exactly at criticality, and that we employ universal counterdiabatic driving \cite{morawetzUniversalCounterdiabaticDriving2025a, finzgarCounterdiabaticDrivingPerformance2025a}, where an alternative variational approach to finding the shortcut should be more robust to the presence of an unknown field.

\begin{figure}
    \centering
    \includegraphics[width=\linewidth]{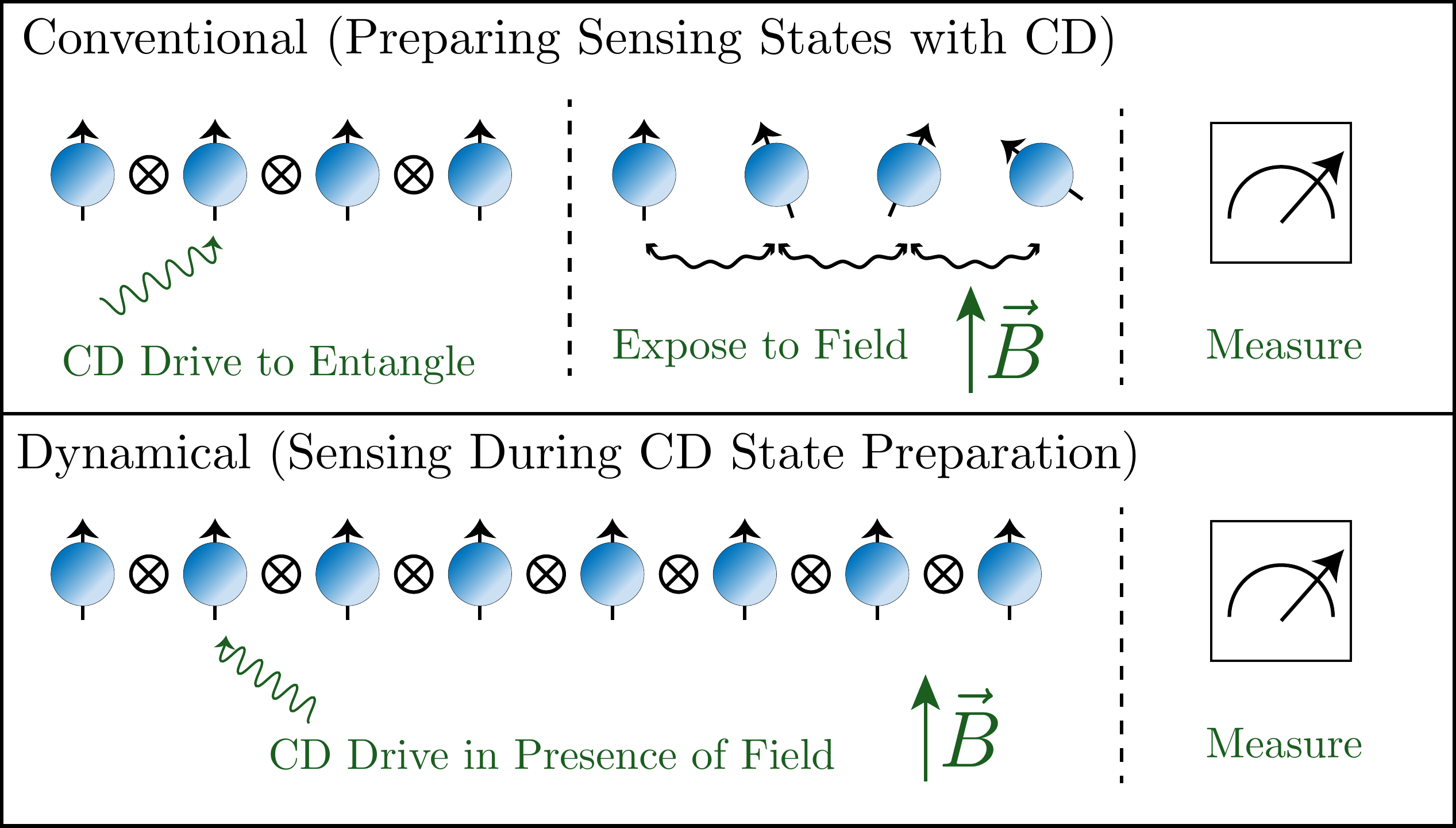}
    \caption{A cartoon illustration of the dynamical sensing procedure. Rather than first try to prepare some entangled state, which is highly fragile, and then use it for sensing, we perform sensing by inferring a parameter from its effect on state preparation. By measuring immediately after the (CD-accelerated) state preparation, the damage done by more rapid decoherence of entangled states is mitigated.}
    \label{fig:cartoon}
\end{figure}

A key advantage to this method is that measurements are made immediately after the protocol is performed. The typical way to perform enhanced sensing is to first prepare some entangled state, and then allow it to undergo free evolution, as in a Ramsey measurement \cite{ramseyMolecularBeamResonance1950}. In this approach, although counterdiabatic driving can accelerate the preparation of these states \cite{hatomuraShortcutsAdiabaticCatstate2018}, they are still subject to the problem of more rapid dephasing. In our approach, however, the sensitivity comes \textit{during} the accelerated preparation. We limit our analysis to measuring immediately after preparation, although some combination of the two has been suggested \cite{hayesMakingMostTime2018}. A cartoon highlighting the differences between our approach and the standard one is shown in Figure \ref{fig:cartoon}.


\prlsection{Universal Counterdiabatic Driving} Our purpose in this work is to demonstrate that a to-be-sensed field may be inferred from how it affects a state preparation protocol. Naturally, this requires a state preparation scheme to begin with. One such class of schemes are adiabatic approaches \cite{kadowakiQuantumAnnealingTransverse1998,farhi2000quantumcomputationadiabaticevolution}, where, provided some external parameter(s) of a system are changed very slowly, the state of a system will follow the instantaneous ground state. If one engineers the Hamiltonian of an isolated system such that a state which is easy to prepare is the initial ground state, and the state of interest is the ground state after the external parameter(s) are changed, one can guarantee that in limit of a very slow change the desired state will be prepared.

In the real world no system is perfectly isolated, and the long times required for adiabatic state preparation are often more than sufficient for noise to destroy the coherence necessary for sensing. In this work, we attempt to mitigate this problem by employing shortcuts to adiabaticity \cite{guery-odelinShortcutsAdiabaticityConcepts2019,duncanTamingQuantumSystems2025a}. This is a broad class of techniques by which adiabatic evolution can be realized faster by modifying the drive of a system.

In particular, we focus on counterdiabatic (CD) driving \cite{demirplakAdiabaticPopulationTransfer2003, demirplakAssistedAdiabaticPassage2005, berryTransitionlessQuantumDriving2009,delcampoAssistedFiniteRateAdiabatic2012,delcampoShortcutsAdiabaticityCounterdiabatic2013,deffnerClassicalQuantumShortcuts2014}. This technique can be understood quite simply: first, consider some adiabatic state preparation scheme where an external parameter $\lambda(t)$ is changed in time. By writing the Schr\"odinger equation in the basis of $\lambda$-dependent eigenstates, it becomes clear that non-adiabatic (diabatic) excitations arise as a result of an effective boost term in this basis, whose strength is proportional to the product of $\dot{\lambda}(t)$ and the adiabatic gauge potential (AGP) \cite{kolodrubetzGeometryNonadiabaticResponse2017}. Excitations will then be suppressed if $\dot{\lambda}(t)$ is sufficiently small (adiabatic), or if the original Hamiltonian is modified so that this boost term is cancelled by adding compensating terms to the driven Hamiltonian (counterdiabatic):
\begin{equation} \label{eq:H_CD}
    H(\lambda) \rightarrow H_{CD}(\lambda) = H(\lambda) + \dot{\lambda} A_\lambda
\end{equation}
\noindent where $A_\lambda$ is the aforementioned AGP. Its definition and relevant properties are given in the End Matter.

Beyond a few simple examples, it is usually very difficult to find the AGP exactly. However, there has been considerable progress in using approximate (local) counterdiabatic driving for state preparation ~\cite{claeysFloquetengineeringCounterdiabaticProtocols2019,passarelliCounterdiabaticDrivingQuantum2020,hartmannPolynomialScalingEnhancement2022,xieVariationalCounterdiabaticDriving2022,cepaiteCounterdiabaticOptimizedLocal2023,schindlerCounterdiabaticDrivingPeriodically2024a,morawetzEfficientPathsLocal2024,petiziolQuantumControlEffective2024a,vanvreumingenGatebasedCounterdiabaticDriving2024a,lawrenceNumericalApproachCalculating2025,gangopadhayCounterdiabaticRouteEntanglement2025a,hsiehLessMoreSubspace2025a,grabaritsFightingExponentiallySmall2026,banerjeePartialReversibilityCounterdiabatic2026}. Recently, a universal approach to counterdiabatic driving was proposed~\cite{morawetzUniversalCounterdiabaticDriving2025a,finzgarCounterdiabaticDrivingPerformance2025a}. There, information about the system enters through the choice of a range of excitation frequencies $[\omega_{min},\omega_{max}]$ to suppress, and finding the CD protocol reduces the the problem of fitting the function $1/\omega$ to a $(2\ell-1)$-th degree odd polynomial in that window (see End Matter for details). In terms of operators, this is an ansatz for the AGP in nested commutators with its order set by $\ell$, which also determines the number of Fourier harmonics required for its Floquet realization ~\cite{claeysFloquetengineeringCounterdiabaticProtocols2019,hatomura2026universaldigitizedcounterdiabaticdriving}.

The primary advantage of this universal approach for use in sensing is that because these protocols attempt to suppress all excitations within a window uniformly, they are much more robust to small uncertainties in the microscopic parameters. As system-specific counterdiabatic protocols are often quite fine-tuned, they should not be expected to perform well in the presence of an unknown parameter \cite{gietkaAdiabaticCriticalQuantum2021}. It is this robustness of the universal approach that makes it well suited for sensing.


\prlsection{Dynamical Sensing Protocol} In the context of enhanced quantum sensing, counterdiabatic driving is usually employed for the purposes of speeding up the preparation of some entangled state \cite{yusteShortcutAdiabaticityInternal2013,campbellShortcutAdiabaticityLipkinMeshkovGlick2015,opatrnyCounterdiabaticDrivingSpin2016,dengisVortexNOONStates2026}. While this shortens the initial state preparation, it does not by itself protect the prepared state from decoherence during the subsequent sensing stage. If the sensitivity instead comes \textit{during} the preparation phase, the preparation might be sped up so that the measurement is made before decoherence becomes a major problem. Exploiting sensitivity during state preparation is not an entirely new idea (see Refs.~[\onlinecite{hayesMakingMostTime2018, gietkaAdiabaticCriticalQuantum2021}]). Our proposed approach combines universal CD driving, which does not require prior knowledge of the parameter being estimated, with a measurement of the field-dependent magnetization generated by finite-time dynamics.

As a demonstration of the proposed approach, we infer the strength of a small longitudinal field present during the preparation of the entangled ground state of the Lipkin-Meshkov-Glick model \cite{LIPKIN1965188,MESHKOV1965199,GLICK1965211}. This spin-$S$ model describes, in particular, $2S$ spin-$\frac{1}{2}$ particles which are all-to-all connected. Such models with a high effective dimensionality are necessary in order to gain a metrological advantage from entanglement \cite{chuStrongQuantumMetrological2023}. Similar models can be realized in experiments, when atoms have an effective interaction mediated by a cavity photon mode which is far detuned from resonance \cite{norciaCavitymediatedCollectiveSpinexchange2018}. Our state preparation is defined by linearly interpolating between two Hamiltonians:

\begin{equation}
    \label{eq:anneal}
    H(\lambda) = \lambda H_0 + (1-\lambda) H_1
\end{equation}

\noindent where the two terms are given by:

\begin{equation}
    \label{eq:LMG}
    H_0 = -\frac{\chi}{2\sqrt{S(S+1)}} S_z^2 - h S_z; \quad H_1 = - g S_x
\end{equation}

\noindent where we set $\chi = g = 1$ to fix units. The prefactor of the nonlinear term is chosen so that $H(\lambda)$ is extensive. One can also rescale Eq. \eqref{eq:anneal} by a factor of $\lambda^{-1}$ (i.e. starting from very large transverse field) so that the to-be-sensed term provides only a constant background, as it would in a real experiment. We tune $\lambda(t)$ from 0 to 1 in a time $\tau$, using the form

\begin{equation}
    \label{eq:lambdat}
    \lambda(t) = \sin^2\left(\frac{\pi}{2} \sin^2\left(\frac{\pi t}{2\tau} \right) \right)
\end{equation}

\noindent which is smooth at $t = 0$ and $t = \tau$ to reduce boundary effect excitations \cite{lidarAdiabaticApproximationExponential2009}. At time $t = \tau$, we immediately measure the observable of interest.


\begin{figure}
    \centering
    \includegraphics[width=\linewidth]{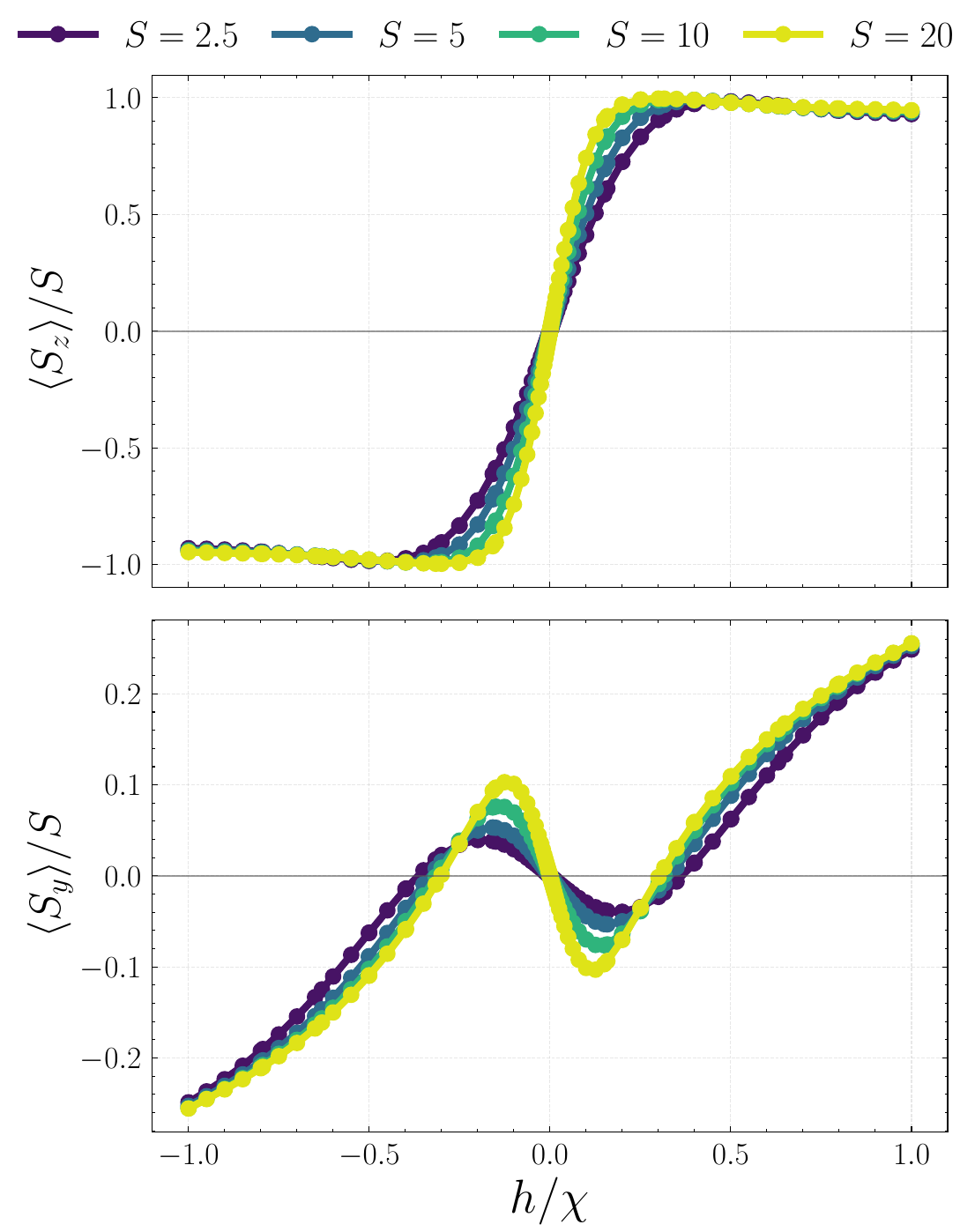}
    \caption{The measured magnetization, normalized by $S$, immediately after performing the dynamical sensing protocol defined by Eq. \eqref{eq:anneal}. Here we choose sample values of $\tau = 1$, $\ell = 6$, $\omega_{max} = 3.7$. The key feature is the response near $h = 0$: sharper growth indicates greater sensitivity to small $h$. This curve can be inverted, at least locally, to estimate $h$ from a measured $S_\alpha$.}
    \label{fig:mag-vs-h}
\end{figure}


\prlsection{Enhanced Sensitivity} To determine if this method provides any actual improvement, we must be able to quantify the sensitivity of our measurement. Theoretically, one can bound the maximum sensitivity achievable by \textit{any} measurement via the Quantum Fisher Information (QFI) \cite{helstromQuantumDetectionEstimation1969, holevoProbabilisticStatisticalAspects2011}. However, the QFI does not directly predict the sensitivity from measuring \textit{accessible} observables. Hence, numerical results and a discussion of the QFI associated with our sensing scheme is relegated to the End Matter. We focus instead on the information which can be extracted from the measurement of spin $\hat{S}_\alpha$ along a direction $\alpha$. Given known parameters: total spin $S$, protocol time $\tau$, the order of the AGP expansion $\ell$, etc., one can compute the mean value $\langle \hat{S}_\alpha\rangle(h)$ via numerical simulation. A typical response of the magnetization to the protocol described above is shown in Figure \ref{fig:mag-vs-h}.

In an experimental setting, the numerically computed response $\langle \hat{S}_\alpha\rangle(h)$ can be inverted locally to define an estimator $h_\alpha(S_\alpha)$ of the unknown field $h$ from a measured value $S_\alpha$. Near $h=0$, this relation can be linearized as $h_\alpha(S_\alpha) =\left.\frac{dh_\alpha}{dS_\alpha}\right|_{S_\alpha=0} S_\alpha
+ O(S_\alpha^2)$. The uncertainty in the inferred field is then related to the spin fluctuations by

\begin{equation}
    \label{eq:sens}
    \Delta h_\alpha=\frac{\Delta S_\alpha}{\vert\partial_h \langle S_\alpha \rangle\vert}
\end{equation}

Thus, in order to improve sensitivity of measuring $h$, we want to maximize the susceptibility $\partial_h \langle S_\alpha \rangle$ at fixed measured-spin fluctuations $\Delta S_\alpha$, which are usually $h$-independent at small $h$. One can already see from Figure \ref{fig:mag-vs-h} that this susceptibility increases with $S$. But, in order to actually benefit from the entanglement induced by the interactions in the system, this susceptibility must increase parametrically faster than $\sqrt{S}$. Thus, our task is to find a measurement scheme under which $\Delta h_\alpha$ decays faster than $1/\sqrt{S}$.

\begin{figure*}[ht]
    \centering
    \includegraphics[width=\linewidth]{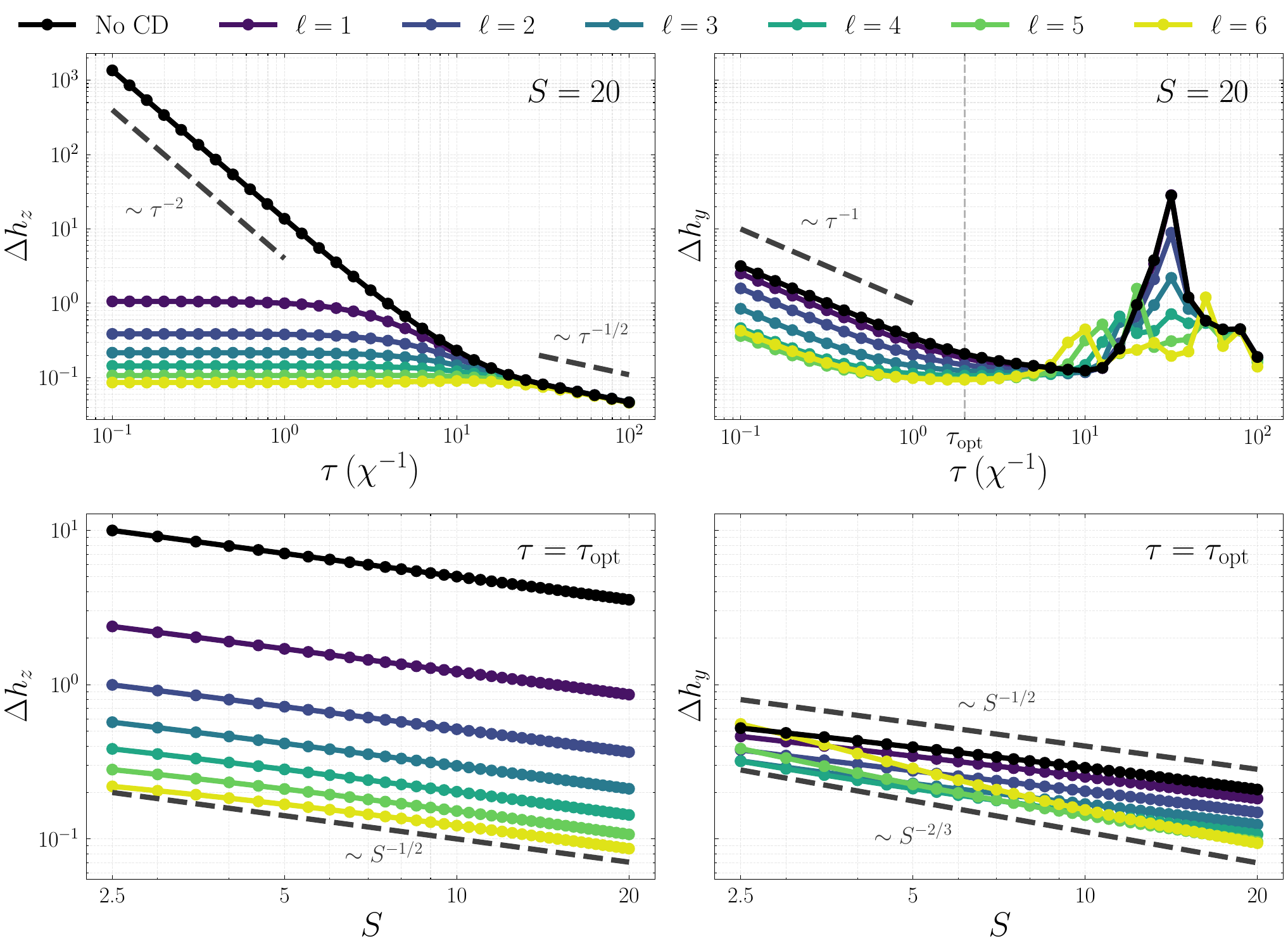}
    \caption{The fluctuations in the estimator of the to-be-sensed parameter $h$, as inferred through measurements of $\hat{S}_z$ (left column) and $\hat{S}_y$ (right column). While $\hat{S}_z$ might seem to be the more natural measurement, it does not scale better with increasing $S$ than performing $S$ measurements independently with $S = 1$. If we instead measure $\hat{S}_y$, we get scaling $\Delta h_y \lesssim S^{-2/3}$ for sufficiently large $\ell$ and $S$. For the universal CD protocol, we use $\omega_{max} = 3.7$ in all plots shown.}
    \label{fig:SNR-results}
\end{figure*}

What observable should we measure? A natural choice might seem to be $\hat{S}_z$. If the system were in equilibrium at the time of measurement, the longitudinal field that we are trying to sense would lead to full polarization along this axis. In this sense, a perfect CD protocol (or an infinitesimally slow adiabatic protocol) will achieve infinite sensitivity. However, a perfect CD protocol requires infinite fine-tuning, which is not possible to achieve without prior knowledge of $h$. In addition to this, the AGP becomes a highly non-local operator near the critical point~\cite{delcampoAssistedFiniteRateAdiabatic2012}. Therefore, away from the very slow limit, imperfections in the counterdiabatic drive will lead to excitations and hence the state will not be perfectly polarized. Nonetheless, we expect some imbalance in the magnetization of the final (non-equilibrium) state, which should lead to a measurable signal.

The sensitivity of a measurement of $\hat{S}_z$ after the dynamical protocol is increased by the presence of CD driving, as shown in the left column of Figure \ref{fig:SNR-results}. It is clear from this that speeding up the dynamical protocol via CD driving leads to greater sensitivity if the protocol is sufficiently fast, and adiabatic sweeps are only comparable at very   long times. However -- as seen in the lower left part of Figure \ref{fig:SNR-results} -- the sensitivity of this measurement scales as $1 / \sqrt{S}$, giving no scaling advantage compared to noninteracting spins.

An alternative, similar to what is done in Ramsey interferometry \cite{ramseyMolecularBeamResonance1950}, is to measure $\hat{S}_y$. There, rather than trying to measure the effect of a field on the equilibrium state, the system is driven out of equilibrium and its effect on the dynamics is measured instead. If our system were truly in equilibrium at $t = \tau$, then we would measure $\langle \hat{S}_y \rangle = 0$. If the protocol is sufficiently slow but non-adiabatic then the outcome of this measurement is very unstable, rapidly oscillating with the protocol time. As can be seen in the top right panel of Figure \ref{fig:SNR-results}, the optimal regime minimizing the measured uncertainty $\Delta h_y$ is achieved at $\tau \sim 1 / \chi$. Notably, for sufficiently large $\ell$ the sensitivity of $\hat S_y$-measurements is beyond the SQL for independent spins. The power-law fit shows that $\Delta h_y \sim S^{-2/3}$ is achievable. Since the scaling improves with $\ell$, we expect that this scaling could be improved even further. We note that as $\ell$ keeps increasing the universal CD protocol may face the so called ``UV problem''~\cite{morawetzUniversalCounterdiabaticDriving2025a, finzgarCounterdiabaticDrivingPerformance2025a} emerging from high-energy excitations. This problem is likely addressed by using a Floquet universal protocol~\cite{hatomura2026universaldigitizedcounterdiabaticdriving}.


\prlsection{Conclusions} In this work, we have proposed a method for achieving high sensitivity during the fast preparation of an entangled state using universal counterdiabatic driving. For the Lipkin--Meshkov--Glick model, we demonstrate beyond-SQL scaling through a simple measurement of the transverse magnetization immediately after the protocol. For the protocols studied, this readout is most sensitive at intermediate durations of order the inverse coupling strength, $\tau\sim1/\chi$. By avoiding a separate sensing stage, our approach may reduce the effects of decoherence associated with long preparation and sensing times~\cite{Boyers_2019} (see discussion in End Matter). The transverse signal giving beyond-SQL sensitivity is encoded in field-dependent deviations from the equilibrium ground state, and vanishes under exact ground-state following. Imperfect CD driving is therefore an essential part of this sensing mechanism. This use of approximate, rather than exact, CD driving parallels proposals for fast cooling~\cite{Villazon_2019} and metastable-state preparation~\cite{Gjonbalaj_2026}. Although we have tested this approach in a specific model, we expect its underlying principles to be applicable more broadly.


\prlsection{Acknowledgments} The authors thank Dries Sels for helpful discussions. This work was supported by the AFOSR Grant FA9550-21-1-0342 and the NSF Grant DMR-2412542. The code used to generate the data and make the plots is available online \footnote{See \url{https://github.com/smorawetz/CD_sensing}}.

\bibliography{bibliography}

@article{foss-feigProgressTrappedIonQuantum2025,
  title = {Progress in {{Trapped-Ion Quantum Simulation}}},
  author = {{Foss-Feig}, Michael and Pagano, Guido and Potter, Andrew C. and Yao, Norman Y.},
  year = 2025,
  month = mar,
  journal = {Annual Review of Condensed Matter Physics},
  volume = {16},
  number = {1},
  pages = {145--172},
  issn = {1947-5454, 1947-5462},
  doi = {10.1146/annurev-conmatphys-032822-045619},
  urldate = {2026-07-13},
  copyright = {http://creativecommons.org/licenses/by/4.0/},
  langid = {english}
}

@article{georgescuQuantumSimulation2014,
  title = {Quantum Simulation},
  author = {Georgescu, I. M. and Ashhab, S. and Nori, Franco},
  year = 2014,
  month = mar,
  journal = {Reviews of Modern Physics},
  volume = {86},
  number = {1},
  pages = {153--185},
  issn = {0034-6861, 1539-0756},
  doi = {10.1103/RevModPhys.86.153},
  urldate = {2026-07-13},
  copyright = {http://link.aps.org/licenses/aps-default-license},
  langid = {english}
}

@article{morgadoQuantumSimulationComputing2021,
  title = {Quantum Simulation and Computing with {{Rydberg-interacting}} Qubits},
  author = {Morgado, M. and Whitlock, S.},
  year = 2021,
  month = jun,
  journal = {AVS Quantum Science},
  volume = {3},
  number = {2},
  pages = {023501},
  issn = {2639-0213},
  doi = {10.1116/5.0036562},
  urldate = {2026-07-13},
  langid = {english}
}

@article{wendinQuantumInformationProcessing2017,
  title = {Quantum Information Processing with Superconducting Circuits: A Review},
  shorttitle = {Quantum Information Processing with Superconducting Circuits},
  author = {Wendin, G},
  year = 2017,
  month = oct,
  journal = {Reports on Progress in Physics},
  volume = {80},
  number = {10},
  pages = {106001},
  issn = {0034-4885, 1361-6633},
  doi = {10.1088/1361-6633/aa7e1a},
  urldate = {2026-07-13},
  langid = {english}
}

@article{degenQuantumSensing2017,
  title = {Quantum Sensing},
  author = {Degen, C. L. and Reinhard, F. and Cappellaro, P.},
  year = 2017,
  month = jul,
  journal = {Reviews of Modern Physics},
  volume = {89},
  number = {3},
  pages = {035002},
  issn = {0034-6861, 1539-0756},
  doi = {10.1103/RevModPhys.89.035002},
  urldate = {2026-06-29},
  copyright = {http://link.aps.org/licenses/aps-default-license},
  langid = {english}
}

@article{cavesQuantummechanicalNoiseInterferometer1981,
  title = {Quantum-Mechanical Noise in an Interferometer},
  author = {Caves, Carlton M.},
  year = 1981,
  month = apr,
  journal = {Physical Review D},
  volume = {23},
  number = {8},
  pages = {1693--1708},
  issn = {0556-2821},
  doi = {10.1103/PhysRevD.23.1693},
  urldate = {2026-07-13},
  copyright = {http://link.aps.org/licenses/aps-default-license},
  langid = {english}
}

@article{kitagawaSqueezedSpinStates1993a,
  title = {Squeezed Spin States},
  author = {Kitagawa, Masahiro and Ueda, Masahito},
  year = 1993,
  month = jun,
  journal = {Physical Review A},
  
  volume = {47},
  number = {6},
  pages = {5138--5143},
  issn = {1050-2947, 1094-1622},
  doi = {10.1103/PhysRevA.47.5138},
  urldate = {2026-07-13},
  copyright = {http://link.aps.org/licenses/aps-default-license},
  langid = {english}
}

@article{winelandSpinSqueezingReduced1992,
  title = {Spin Squeezing and Reduced Quantum Noise in Spectroscopy},
  author = {Wineland, D. J. and Bollinger, J. J. and Itano, W. M. and Moore, F. L. and Heinzen, D. J.},
  year = 1992,
  month = dec,
  journal = {Physical Review A},
  volume = {46},
  number = {11},
  pages = {R6797-R6800},
  issn = {1050-2947, 1094-1622},
  doi = {10.1103/PhysRevA.46.R6797},
  urldate = {2026-06-05},
  copyright = {http://link.aps.org/licenses/aps-default-license},
  langid = {english}
}

@article{boixoGeneralizedLimitsSingleParameter2007,
  title = {Generalized {{Limits}} for {{Single-Parameter Quantum Estimation}}},
  author = {Boixo, Sergio and Flammia, Steven T. and Caves, Carlton M. and Geremia, Jm},
  year = 2007,
  month = feb,
  journal = {Physical Review Letters},
  volume = {98},
  number = {9},
  pages = {090401},
  issn = {0031-9007, 1079-7114},
  doi = {10.1103/PhysRevLett.98.090401},
  urldate = {2026-07-01},
  copyright = {http://link.aps.
  org/licenses/aps-default-license},
  langid = {english}
}

@article{giovannettiQuantumEnhancedMeasurementsBeating2004,
  title = {Quantum-{{Enhanced Measurements}}: {{Beating}} the {{Standard Quantum Limit}}},
  shorttitle = {Quantum-{{Enhanced Measurements}}},
  author = {Giovannetti, Vittorio and Lloyd, Seth and Maccone, Lorenzo},
  year = 2004,
  month = nov,
  journal = {Science},
  volume = {306},
  number = {5700},
  pages = {1330--1336},
  issn = {0036-8075, 1095-9203},
  doi = {10.1126/science.1104149},
  urldate = {2026-07-13},
  langid = {english}
}

@article{kesslerHeisenbergLimitedAtomClocks2014,
  title = {Heisenberg-{{Limited Atom Clocks Based}} on {{Entangled Qubits}}},
  author = {Kessler, E. M. and K{\'o}m{\'a}r, P. and Bishof, M. and Jiang, L. and S{\o}rensen, A. S. and Ye, J. and Lukin, M. D.},
  year = 2014,
  month = may,
  journal = {Physical Review Letters},
  volume = {112},
  number = {19},
  pages = {190403},
  issn = {0031-9007, 1079-7114},
  doi = {10.1103/PhysRevLett.112.190403},
  urldate = {2026-07-14},
  copyright = {http://link.aps.org/licenses/aps-default-license},
  langid = {english}
}

@misc{dengisVortexNOONStates2026,
  title = {Vortex {{NOON}} States for Rotation Sensing},
  author = {Dengis, Simon and Dupont, Nathan and Schlagheck, Peter and Goldman, Nathan},
  year = 2026,
  month = jun,
  number = {arXiv:2606.29509},
  eprint = {2606.29509},
  primaryclass = {cond-mat.quant-gas},
  publisher = {arXiv},
  doi = {10.48550/arXiv.2606.29509},
  urldate = {2026-06-30},
  archiveprefix = {arXiv},
  langid = {english}
}

@article{omranGenerationManipulationSchrodinger2019,
  title = {Generation and Manipulation of {{Schr\"odinger}} Cat States in {{Rydberg}} Atom Arrays},
  author = {Omran, A. and Levine, H. and Keesling, A. and Semeghini, G. and Wang, T. T. and Ebadi, S. and Bernien, H. and Zibrov, A. S. and Pichler, H. and Choi, S. and Cui, J. and Rossignolo, M. and Rembold, P. and Montangero, S. and Calarco, T. and Endres, M. and Greiner, M. and Vuleti{\'c}, V. and Lukin, M. D.},
  year = 2019,
  month = aug,
  journal = {Science},
  volume = {365},
  number = {6453},
  pages = {570--574},
  issn = {0036-8075, 1095-9203},
  doi = {10.1126/science.aax9743},
  urldate = {2026-07-14},
  langid = {english}
}

@article{friisObservationEntangledStates2018a,
  title = {Observation of {{Entangled States}} of a {{Fully Controlled}} 20-{{Qubit System}}},
  author = {Friis, Nicolai and Marty, Oliver and Maier, Christine and Hempel, Cornelius and Holz{\"a}pfel, Milan and Jurcevic, Petar and Plenio, Martin B. and Huber, Marcus and Roos, Christian and Blatt, Rainer and Lanyon, Ben},
  year = 2018,
  month = apr,
  journal = {Physical Review X},
  volume = {8},
  number = {2},
  pages = {021012},
  issn = {2160-3308},
  doi = {10.1103/PhysRevX.8.021012},
  urldate = {2026-07-14},
  langid = {english}
}

@article{bohmannEntanglementPhaseProperties2015,
  title = {Entanglement and Phase Properties of Noisy {{NOON}} States},
  author = {Bohmann, M. and Sperling, J. and Vogel, W.},
  year = 2015,
  month = apr,
  journal = {Physical Review A},
  volume = {91},
  number = {4},
  pages = {042332},
  issn = {1050-2947, 1094-1622},
  doi = {10.1103/PhysRevA.91.042332},
  urldate = {2026-07-14},
  copyright = {http://link.aps.org/licenses/aps-default-license},
  langid = {english}
}

@article{huelgaImprovementFrequencyStandards1997a,
  title = {Improvement of {{Frequency Standards}} with {{Quantum Entanglement}}},
  author = {Huelga, S. F. and Macchiavello, C. and Pellizzari, T. and Ekert, A. K. and Plenio, M. B. and Cirac, J. I.},
  year = 1997,
  month = nov,
  journal = {Physical Review Letters},
  volume = {79},
  number = {20},
  pages = {3865--3868},
  issn = {0031-9007, 1079-7114},
  doi = {10.1103/PhysRevLett.79.3865},
  urldate = {2026-07-13},
  copyright = {http://link.aps.org/licenses/aps-default-license},
  langid = {english}
}

@article{bollingerOptimalFrequencyMeasurements1996,
  title = {Optimal Frequency Measurements with Maximally Correlated States},
  author = {Bollinger, J. J . and Itano, Wayne M. and Wineland, D. J. and Heinzen, D. J.},
  year = 1996,
  month = dec,
  journal = {Physical Review A},
  volume = {54},
  number = {6},
  pages = {R4649-R4652},
  issn = {1050-2947, 1094-1622},
  doi = {10.1103/PhysRevA.54.R4649},
  urldate = {2026-07-14},
  copyright = {http://link.aps.org/licenses/aps-default-license},
  langid = {english}
}

@article{greenbergerBellsTheoremInequalities1990,
  title = {Bell's Theorem without Inequalities},
  author = {Greenberger, Daniel M. and Horne, Michael A. and Shimony, Abner and Zeilinger, Anton},
  year = 1990,
  month = dec,
  journal = {American Journal of Physics},
  volume = {58},
  number = {12},
  pages = {1131--1143},
  issn = {0002-9505, 1943-2909},
  doi = {10.1119/1.16243},
  urldate = {2026-07-14},
  langid = {english}
}

@article{kokCreationLargephotonnumberPath2002,
  title = {Creation of Large-Photon-Number Path Entanglement Conditioned on Photodetection},
  author = {Kok, Pieter and Lee, Hwang and Dowling, Jonathan P.},
  year = 2002,
  month = apr,
  journal = {Physical Review A},
  volume = {65},
  number = {5},
  pages = {052104},
  issn = {1050-2947, 1094-1622},
  doi = {10.1103/PhysRevA.65.052104},
  urldate = {2026-07-14},
  copyright = {http://link.aps.org/licenses/aps-default-license},
  langid = {english}
}

@article{dowlingQuantumOpticalMetrology2008,
  title = {Quantum Optical Metrology -- the Lowdown on High-{{N00N}} States},
  author = {Dowling, Jonathan P.},
  year = 2008,
  month = mar,
  journal = {Contemporary Physics},
  volume = {49},
  number = {2},
  pages = {125--143},
  issn = {0010-7514, 1366-5812},
  doi = {10.1080/00107510802091298},
  urldate = {2026-07-14},
  langid = {english}
}

@article{mitchellSuperresolvingPhaseMeasurements2004,
  title = {Super-Resolving Phase Measurements with a Multiphoton Entangled State},
  author = {Mitchell, M. W. and Lundeen, J. S. and Steinberg, A. M.},
  year = 2004,
  month = may,
  journal = {Nature},
  volume = {429},
  number = {6988},
  pages = {161--164},
  issn = {0028-0836, 1476-4687},
  doi = {10.1038/nature02493},
  urldate = {2026-07-14},
  copyright = {http://www.springer.com/tdm},
  langid = {english}
}

@article{helstromQuantumDetectionEstimation1969,
  title = {Quantum Detection and Estimation Theory},
  author = {Helstrom, Carl W.},
  year = 1969,
  month = jun,
  journal = {Journal of Statistical Physics},
  volume = {1},
  number = {2},
  pages = {231--252},
  issn = {1572-9613},
  doi = {10.1007/BF01007479}
}

@book{holevoProbabilisticStatisticalAspects2011,
  title = {Probabilistic and {{Statistical Aspects}} of {{Quantum Theory}}},
  author = {Holevo, Alexander Semenovich},
  year = 2011,
  series = {Publications of the {{Scuola Normale Superiore}}},
  publisher = {Springer Basel Springer e-books},
  address = {Pisa},
  isbn = {978-88-7642-378-9},
  langid = {english},
  lccn = {530.133}
}

@article{braunsteinStatisticalDistanceGeometry1994,
  title = {Statistical Distance and the Geometry of Quantum States},
  author = {Braunstein, Samuel L. and Caves, Carlton M.},
  year = 1994,
  month = may,
  journal = {Physical Review Letters},
  volume = {72},
  number = {22},
  pages = {3439--3443},
  issn = {0031-9007},
  doi = {10.1103/PhysRevLett.72.3439},
  urldate = {2026-07-02},
  copyright = {http://link.aps.org/licenses/aps-default-license},
  langid = {english}
}

@article{frerotQuantumCriticalMetrology2018,
  title = {Quantum {{Critical Metrology}}},
  author = {Fr{\'e}rot, Ir{\'e}n{\'e}e and Roscilde, Tommaso},
  year = 2018,
  month = jul,
  journal = {Physical Review Letters},
  volume = {121},
  number = {2},
  pages = {020402},
  issn = {0031-9007, 1079-7114},
  doi = {10.1103/PhysRevLett.121.020402},
  urldate = {2026-08-05},
  langid = {english}
}

@article{mihailescuCriticalQuantumSensing2026,
  title = {Critical {{Quantum Sensing}}: {{A Tutorial}} on {{Parameter Estimation Near Quantum Phase Transitions}}},
  shorttitle = {Critical {{Quantum Sensing}}},
  author = {Mihailescu, George and Alushi, Uesli and Di Candia, Roberto and Felicetti, Simone and Gietka, Karol},
  year = 2026,
  month = jun,
  journal = {PRX Quantum},
  volume = {7},
  number = {2},
  pages = {020201},
  issn = {2691-3399},
  doi = {10.1103/v7mf-yh8n},
  urldate = {2026-07-14},
  langid = {english}
}

@article{zanardiQuantumCriticalityResource2008,
  title = {Quantum Criticality as a Resource for Quantum Estimation},
  author = {Zanardi, Paolo and Paris, Matteo G. A. and Campos Venuti, Lorenzo},
  year = 2008,
  month = oct,
  journal = {Physical Review A},
  volume = {78},
  number = {4},
  pages = {042105},
  issn = {1050-2947, 1094-1622},
  doi = {10.1103/PhysRevA.78.042105},
  urldate = {2026-05-20},
  copyright = {http://link.aps.org/licenses/aps-default-license},
  langid = {english}
}

@article{ramsLimitsCriticalityBasedQuantum2018,
  title = {At the {{Limits}} of {{Criticality-Based Quantum Metrology}}: {{Apparent Super-Heisenberg Scaling Revisited}}},
  shorttitle = {At the {{Limits}} of {{Criticality-Based Quantum Metrology}}},
  author = {Rams, Marek M. and Sierant, Piotr and Dutta, Omyoti and Horodecki, Pawe{\l} and Zakrzewski, Jakub},
  year = 2018,
  month = apr,
  journal = {Physical Review X},
  volume = {8},
  number = {2},
  pages = {021022},
  issn = {2160-3308},
  doi = {10.1103/PhysRevX.8.021022},
  urldate = {2026-08-05},
  langid = {english}
}

@article{gietkaAdiabaticCriticalQuantum2021,
  title = {Adiabatic Critical Quantum Metrology Cannot Reach the {{Heisenberg}} Limit Even When Shortcuts to Adiabaticity Are Applied},
  author = {Gietka, Karol and Metz, Friederike and Keller, Tim and Li, Jing},
  year = 2021,
  month = jul,
  journal = {Quantum},
  volume = {5},
  eprint = {2103.12939},
  primaryclass = {quant-ph},
  pages = {489},
  issn = {2521-327X},
  doi = {10.22331/q-2021-07-01-489},
  urldate = {2026-07-01},
  archiveprefix = {arXiv},
  langid = {english}
}

@article{finzgarCounterdiabaticDrivingPerformance2025a,
  title = {Counterdiabatic {{Driving}} with {{Performance Guarantees}}},
  author = {Fin{\v z}gar, Jernej Rudi and Notarnicola, Simone and Cain, Madelyn and Lukin, Mikhail D. and Sels, Dries},
  year = 2025,
  month = oct,
  journal = {Physical Review Letters},
  volume = {135},
  number = {18},
  pages = {180602},
  issn = {0031-9007, 1079-7114},
  doi = {10.1103/pqhl-nbtk},
  urldate = {2026-08-06},
  langid = {english}
}

@article{morawetzUniversalCounterdiabaticDriving2025a,
  title = {Universal {{Counterdiabatic Driving}} in {{Krylov Space}}},
  author = {Morawetz, Stewart and Polkovnikov, Anatoli},
  year = 2025,
  month = oct,
  journal = {PRX Quantum},
  volume = {6},
  number = {4},
  pages = {040320},
  issn = {2691-3399},
  doi = {10.1103/wbbs-s8fs},
  urldate = {2026-08-06},
  langid = {english}
}

@article{ramseyMolecularBeamResonance1950,
  title = {A {{Molecular Beam Resonance Method}} with {{Separated Oscillating Fields}}},
  author = {Ramsey, Norman F.},
  year = 1950,
  month = jun,
  journal = {Physical Review},
  volume = {78},
  pages = {695--699},
  publisher = {APS},
  issn = {1536-6065},
  doi = {10.1103/PhysRev.78.695},
  urldate = {2026-08-06}
}

@article{hatomuraShortcutsAdiabaticCatstate2018,
  title = {Shortcuts to Adiabatic Cat-State Generation in Bosonic {{Josephson}} Junctions},
  author = {Hatomura, Takuya},
  year = 2018,
  month = jan,
  journal = {New Journal of Physics},
  volume = {20},
  number = {1},
  pages = {015010},
  issn = {1367-2630},
  doi = {10.1088/1367-2630/aaa117},
  urldate = {2026-08-07}
}

@article{hayesMakingMostTime2018,
  title = {Making the Most of Time in Quantum Metrology: Concurrent State Preparation and Sensing},
  shorttitle = {Making the Most of Time in Quantum Metrology},
  author = {Hayes, Anthony J and Dooley, Shane and Munro, William J and Nemoto, Kae and Dunningham, Jacob},
  year = 2018,
  month = jul,
  journal = {Quantum Science and Technology},
  volume = {3},
  number = {3},
  pages = {035007},
  issn = {2058-9565},
  doi = {10.1088/2058-9565/aac30b},
  urldate = {2026-07-15},
  langid = {english}
}

@misc{farhi2000quantumcomputationadiabaticevolution,
  title = {Quantum Computation by Adiabatic Evolution},
  author = {Farhi, Edward and Goldstone, Jeffrey and Gutmann, Sam and Sipser, Michael},
  year = 2000,
  eprint = {quant-ph/0001106},
  archiveprefix = {arXiv}
}

@article{kadowakiQuantumAnnealingTransverse1998,
  title = {Quantum Annealing in the Transverse {{Ising}} Model},
  author = {Kadowaki, Tadashi and Nishimori, Hidetoshi},
  year = 1998,
  month = nov,
  journal = {Physical Review E},
  volume = {58},
  number = {5},
  pages = {5355--5363},
  issn = {1063-651X, 1095-3787},
  doi = {10.1103/PhysRevE.58.5355},
  urldate = {2026-08-25},
  copyright = {http://link.aps.org/licenses/aps-default-license},
  langid = {english}
}

@article{berryTransitionlessQuantumDriving2009,
  title = {Transitionless Quantum Driving},
  author = {Berry, M V},
  year = 2009,
  month = sep,
  journal = {Journal of Physics A: Mathematical and Theoretical},
  volume = {42},
  number = {36},
  pages = {365303},
  issn = {1751-8113, 1751-8121},
  doi = {10.1088/1751-8113/42/36/365303},
  urldate = {2022-05-16},
  langid = {english}
}

@article{demirplakAdiabaticPopulationTransfer2003,
  title = {Adiabatic {{Population Transfer}} with {{Control Fields}}},
  author = {Demirplak, Mustafa and Rice, Stuart A.},
  year = 2003,
  month = nov,
  journal = {The Journal of Physical Chemistry A},
  volume = {107},
  number = {46},
  pages = {9937--9945},
  issn = {1089-5639, 1520-5215},
  doi = {10.1021/jp030708a},
  urldate = {2024-10-30},
  langid = {english}
}

@article{demirplakAssistedAdiabaticPassage2005,
  title = {Assisted {{Adiabatic Passage Revisited}}},
  author = {Demirplak, Mustafa and Rice, Stuart A.},
  year = 2005,
  month = apr,
  journal = {The Journal of Physical Chemistry B},
  volume = {109},
  number = {14},
  pages = {6838--6844},
  issn = {1520-6106, 1520-5207},
  doi = {10.1021/jp040647w},
  urldate = {2024-10-30},
  langid = {english}
}

@article{guery-odelinShortcutsAdiabaticityConcepts2019,
  title = {Shortcuts to Adiabaticity: {{Concepts}}, Methods, and Applications},
  shorttitle = {Shortcuts to Adiabaticity},
  author = {{Gu{\'e}ry-Odelin}, D. and Ruschhaupt, A. and Kiely, A. and Torrontegui, E. and {Mart{\'i}nez-Garaot}, S. and Muga, J. G.},
  year = 2019,
  month = oct,
  journal = {Reviews of Modern Physics},
  volume = {91},
  number = {4},
  pages = {045001},
  issn = {0034-6861, 1539-0756},
  doi = {10.1103/RevModPhys.91.045001},
  urldate = {2025-12-19},
  langid = {english}
}

@article{kolodrubetzGeometryNonadiabaticResponse2017,
  title = {Geometry and Non-Adiabatic Response in Quantum and Classical Systems},
  author = {Kolodrubetz, Michael and Sels, Dries and Mehta, Pankaj and Polkovnikov, Anatoli},
  year = 2017,
  month = jun,
  journal = {Physics Reports},
  volume = {697},
  eprint = {1602.01062},
  primaryclass = {cond-mat},
  pages = {1--87},
  issn = {03701573},
  doi = {10.1016/j.physrep.2017.07.001},
  urldate = {2023-03-03},
  archiveprefix = {arXiv}
}

@article{deffnerClassicalQuantumShortcuts2014,
  title = {Classical and {{Quantum Shortcuts}} to {{Adiabaticity}} for {{Scale-Invariant Driving}}},
  author = {Deffner, Sebastian and Jarzynski, Christopher and {del Campo}, Adolfo},
  year = 2014,
  month = apr,
  journal = {Physical Review X},
  volume = {4},
  number = {2},
  pages = {021013},
  issn = {2160-3308},
  doi = {10.1103/PhysRevX.4.021013},
  urldate = {2023-03-03},
  langid = {english}
}

@article{delcampoAssistedFiniteRateAdiabatic2012,
  title = {Assisted {{Finite-Rate Adiabatic Passage Across}} a {{Quantum Critical Point}}: {{Exact Solution}} for the {{Quantum Ising Model}}},
  shorttitle = {Assisted {{Finite-Rate Adiabatic Passage Across}} a {{Quantum Critical Point}}},
  author = {Del Campo, Adolfo and Rams, Marek M. and Zurek, Wojciech H.},
  year = 2012,
  month = sep,
  journal = {Physical Review Letters},
  volume = {109},
  number = {11},
  pages = {115703},
  issn = {0031-9007, 1079-7114},
  doi = {10.1103/PhysRevLett.109.115703},
  urldate = {2026-08-26},
  copyright = {http://link.aps.org/licenses/aps-default-license},
  langid = {english}
}

@article{delcampoShortcutsAdiabaticityCounterdiabatic2013,
  title = {Shortcuts to {{Adiabaticity}} by {{Counterdiabatic Driving}}},
  author = {{del Campo}, Adolfo},
  year = 2013,
  month = sep,
  journal = {Physical Review Letters},
  volume = {111},
  number = {10},
  pages = {100502},
  publisher = {American Physical Society},
  doi = {10.1103/PhysRevLett.111.100502},
  urldate = {2024-10-30}
}

@article{selsMinimizingIrreversibleLosses2017,
  title = {Minimizing Irreversible Losses in Quantum Systems by Local Counterdiabatic Driving},
  author = {Sels, Dries and Polkovnikov, Anatoli},
  year = 2017,
  month = may,
  journal = {Proceedings of the National Academy of Sciences},
  volume = {114},
  number = {20},
  issn = {0027-8424, 1091-6490},
  doi = {10.1073/pnas.1619826114},
  urldate = {2022-05-29},
  langid = {english}
}

@article{duncanTamingQuantumSystems2025a,
  title = {Taming {{Quantum Systems}}: {{A Tutorial}} for {{Using Shortcuts-To-Adiabaticity}}, {{Quantum Optimal Control}}, and {{Reinforcement Learning}}},
  shorttitle = {Taming {{Quantum Systems}}},
  author = {Duncan, Callum W. and Poggi, Pablo M. and Bukov, Marin and Zinner, Nikolaj Thomas and Campbell, Steve},
  year = 2025,
  month = oct,
  journal = {PRX Quantum},
  volume = {6},
  number = {4},
  pages = {040201},
  issn = {2691-3399},
  doi = {10.1103/j8c7-v2hd},
  urldate = {2026-08-26},
  langid = {english}
}

@article{claeysFloquetengineeringCounterdiabaticProtocols2019,
  title = {Floquet-Engineering Counterdiabatic Protocols in Quantum Many-Body Systems},
  author = {Claeys, Pieter W. and Pandey, Mohit and Sels, Dries and Polkovnikov, Anatoli},
  year = 2019,
  month = aug,
  journal = {Physical Review Letters},
  volume = {123},
  number = {9},
  eprint = {1904.03209},
  pages = {090602},
  issn = {0031-9007, 1079-7114},
  doi = {10.1103/PhysRevLett.123.090602},
  urldate = {2022-03-19},
  archiveprefix = {arXiv}
}

@article{lawrenceNumericalApproachCalculating2025,
  title = {A Numerical Approach for Calculating Exact Non-Adiabatic Terms in Quantum Dynamics},
  author = {Lawrence, Ewen and Schmid, Sebastian F J and {\v C}epait{\.e}, Ieva and Kirton, Peter and Duncan, Callum W},
  year = 2025,
  month = jan,
  journal = {SciPost Physics},
  volume = {18},
  number = {1},
  pages = {014},
  issn = {2542-4653},
  doi = {10.21468/SciPostPhys.18.1.014},
  urldate = {2026-08-26}
}

@misc{hsiehLessMoreSubspace2025a,
  title = {Less Is More: Subspace Reduction for Counterdiabatic Driving of {{Rydberg}} Atom Arrays},
  shorttitle = {Less Is More},
  author = {Hsieh, Wen Ting and Sels, Dries},
  year = 2025,
  month = dec,
  number = {arXiv:2512.04494},
  eprint = {2512.04494},
  primaryclass = {quant-ph},
  publisher = {arXiv},
  doi = {10.48550/arXiv.2512.04494},
  urldate = {2026-08-26},
  archiveprefix = {arXiv}
}

@article{vanvreumingenGatebasedCounterdiabaticDriving2024a,
  title = {Gate-Based Counterdiabatic Driving with Complexity Guarantees},
  author = {Van Vreumingen, Dyon},
  year = 2024,
  month = nov,
  journal = {Physical Review A},
  volume = {110},
  number = {5},
  pages = {052419},
  issn = {2469-9926, 2469-9934},
  doi = {10.1103/PhysRevA.110.052419},
  urldate = {2026-08-26},
  langid = {english}
}

@article{grabaritsFightingExponentiallySmall2026,
  title = {Fighting {{Exponentially Small Gaps}} by {{Counterdiabatic Driving}}},
  author = {Grabarits, Andr{\'a}s and Balducci, Federico and Del Campo, Adolfo},
  year = 2026,
  month = feb,
  journal = {PRX Quantum},
  volume = {7},
  number = {1},
  pages = {010322},
  issn = {2691-3399},
  doi = {10.1103/tgzt-dy3h},
  urldate = {2026-08-26},
  langid = {english}
}

@article{morawetzEfficientPathsLocal2024,
  title = {Efficient Paths for Local Counterdiabatic Driving},
  author = {Morawetz, Stewart and Polkovnikov, Anatoli},
  year = 2024,
  month = jul,
  journal = {Physical Review B},
  volume = {110},
  number = {2},
  pages = {024304},
  publisher = {American Physical Society},
  doi = {10.1103/PhysRevB.110.024304},
  urldate = {2024-10-30}
}

@misc{banerjeePartialReversibilityCounterdiabatic2026,
  title = {Partial {{Reversibility}} and {{Counterdiabatic Driving}} in {{Nearly Integrable Systems}}},
  author = {Banerjee, Rohan and Khamnei, Shahyad and Polkovnikov, Anatoli and Morawetz, Stewart},
  year = 2026,
  month = jun,
  number = {arXiv:2602.22317},
  eprint = {2602.22317},
  primaryclass = {quant-ph},
  publisher = {arXiv},
  doi = {10.48550/arXiv.2602.22317},
  urldate = {2026-08-26},
  archiveprefix = {arXiv}
}

@article{gangopadhayCounterdiabaticRouteEntanglement2025a,
  title = {Counterdiabatic {{Route}} to {{Entanglement Steering}} and {{Dynamical Freezing}} in the {{Floquet Lipkin-Meshkov-Glick Model}}},
  author = {Gangopadhay, Nakshatra and Choudhury, Sayan},
  year = 2025,
  month = jul,
  journal = {Physical Review Letters},
  volume = {135},
  number = {2},
  pages = {020407},
  issn = {0031-9007, 1079-7114},
  doi = {10.1103/bzcf-gm89},
  urldate = {2026-08-26},
  langid = {english}
}

@article{schindlerCounterdiabaticDrivingPeriodically2024a,
  title = {Counterdiabatic {{Driving}} for {{Periodically Driven Systems}}},
  author = {Schindler, Paul M. and Bukov, Marin},
  year = 2024,
  month = sep,
  journal = {Physical Review Letters},
  volume = {133},
  number = {12},
  pages = {123402},
  issn = {0031-9007, 1079-7114},
  doi = {10.1103/PhysRevLett.133.123402},
  urldate = {2026-08-26},
  langid = {english}
}

@article{cepaiteCounterdiabaticOptimizedLocal2023,
  title = {Counterdiabatic {{Optimized Local Driving}}},
  author = {{\v C}epait{\.e}, Ieva and Polkovnikov, Anatoli and Daley, Andrew J. and Duncan, Callum W.},
  year = 2023,
  month = jan,
  journal = {PRX Quantum},
  volume = {4},
  number = {1},
  pages = {010312},
  issn = {2691-3399},
  doi = {10.1103/PRXQuantum.4.010312},
  urldate = {2024-10-30},
  langid = {english}
}

@article{petiziolQuantumControlEffective2024a,
  title = {Quantum Control by Effective Counterdiabatic Driving},
  author = {Petiziol, Francesco and Mintert, Florian and Wimberger, Sandro},
  year = 2024,
  month = jan,
  journal = {Europhysics Letters},
  volume = {145},
  number = {1},
  pages = {15001},
  issn = {0295-5075, 1286-4854},
  doi = {10.1209/0295-5075/ad19e3},
  urldate = {2026-08-26}
}

@article{hartmannPolynomialScalingEnhancement2022,
  title = {Polynomial Scaling Enhancement in the Ground-State Preparation of {{Ising}} Spin Models via Counterdiabatic Driving},
  author = {Hartmann, Andreas and Mbeng, Glen Bigan and Lechner, Wolfgang},
  year = 2022,
  month = feb,
  journal = {Physical Review A},
  volume = {105},
  number = {2},
  pages = {022614},
  issn = {2469-9926, 2469-9934},
  doi = {10.1103/PhysRevA.105.022614},
  urldate = {2026-08-26},
  langid = {english}
}

@article{passarelliCounterdiabaticDrivingQuantum2020,
  title = {Counterdiabatic Driving in the Quantum Annealing of the p -Spin Model: {{A}} Variational Approach},
  shorttitle = {Counterdiabatic Driving in the Quantum Annealing of the p -Spin Model},
  author = {Passarelli, G. and Cataudella, V. and Fazio, R. and Lucignano, P.},
  year = 2020,
  month = mar,
  journal = {Physical Review Research},
  volume = {2},
  number = {1},
  pages = {013283},
  issn = {2643-1564},
  doi = {10.1103/PhysRevResearch.2.013283},
  urldate = {2026-08-26},
  langid = {english}
}

@article{xieVariationalCounterdiabaticDriving2022,
  title = {Variational Counterdiabatic Driving of the {{Hubbard}} Model for Ground-State Preparation},
  author = {Xie, Qing and Seki, Kazuhiro and Yunoki, Seiji},
  year = 2022,
  month = oct,
  journal = {Physical Review B},
  volume = {106},
  number = {15},
  pages = {155153},
  issn = {2469-9950, 2469-9969},
  doi = {10.1103/PhysRevB.106.155153},
  urldate = {2026-08-26},
  langid = {english}
}

@article{takahashiShortcutsAdiabaticityKrylov2024,
  title = {Shortcuts to {{Adiabaticity}} in {{Krylov Space}}},
  author = {Takahashi, Kazutaka and Del Campo, Adolfo},
  year = 2024,
  month = feb,
  journal = {Physical Review X},
  volume = {14},
  number = {1},
  pages = {011032},
  issn = {2160-3308},
  doi = {10.1103/PhysRevX.14.011032},
  urldate = {2024-11-26},
  langid = {english}
}

@misc{bhattacharjeeLanczosApproachAdiabatic2023,
  title = {A {{Lanczos}} Approach to the {{Adiabatic Gauge Potential}}},
  author = {Bhattacharjee, Budhaditya},
  year = 2023,
  month = feb,
  number = {arXiv:2302.07228},
  eprint = {2302.07228},
  publisher = {arXiv},
  doi = {10.48550/arXiv.2302.07228},
  urldate = {2024-11-26},
  archiveprefix = {arXiv}
}

@article{campbellShortcutAdiabaticityLipkinMeshkovGlick2015,
  title = {Shortcut to {{Adiabaticity}} in the {{Lipkin-Meshkov-Glick Model}}},
  author = {Campbell, Steve and De Chiara, Gabriele and Paternostro, Mauro and Palma, G. Massimo and Fazio, Rosario},
  year = 2015,
  month = may,
  journal = {Physical Review Letters},
  volume = {114},
  number = {17},
  pages = {177206},
  issn = {0031-9007, 1079-7114},
  doi = {10.1103/PhysRevLett.114.177206},
  urldate = {2026-09-01},
  copyright = {http://link.aps.org/licenses/aps-default-license},
  langid = {english}
}

@article{opatrnyCounterdiabaticDrivingSpin2016,
  title = {Counterdiabatic Driving in Spin Squeezing and {{Dicke-state}} Preparation},
  author = {Opatrn{\'y}, Tom{\'a}{\v s} and Saberi, Hamed and Brion, Etienne and M{\o}lmer, Klaus},
  year = 2016,
  month = feb,
  journal = {Physical Review A},
  volume = {93},
  number = {2},
  pages = {023815},
  issn = {2469-9926, 2469-9934},
  doi = {10.1103/PhysRevA.93.023815},
  urldate = {2026-09-01},
  copyright = {http://link.aps.org/licenses/aps-default-license},
  langid = {english}
}

@article{yusteShortcutAdiabaticityInternal2013,
  title = {Shortcut to Adiabaticity in Internal Bosonic {{Josephson}} Junctions},
  author = {Yuste, A. and {Juli{\'a}-D{\'i}az}, B. and Torrontegui, E. and Martorell, J. and Muga, J. G. and Polls, A.},
  year = 2013,
  month = oct,
  journal = {Physical Review A},
  volume = {88},
  number = {4},
  pages = {043647},
  issn = {1050-2947, 1094-1622},
  doi = {10.1103/PhysRevA.88.043647},
  urldate = {2026-09-02},
  copyright = {http://link.aps.org/licenses/aps-default-license},
  langid = {english}
}

@article{GLICK1965211,
  title = {Validity of Many-Body Approximation Methods for a Solvable Model: ({{III}}). {{Diagram}} Summations},
  author = {Glick, A.J. and Lipkin, H.J. and Meshkov, N.},
  year = 1965,
  journal = {Nuclear Physics},
  volume = {62},
  number = {2},
  pages = {211--224},
  issn = {0029-5582},
  doi = {10.1016/0029-5582(65)90864-3}
}

@article{LIPKIN1965188,
  title = {Validity of Many-Body Approximation Methods for a Solvable Model: ({{I}}). {{Exact}} Solutions and Perturbation Theory},
  author = {Lipkin, H.J. and Meshkov, N. and Glick, A.J.},
  year = 1965,
  journal = {Nuclear Physics},
  volume = {62},
  number = {2},
  pages = {188--198},
  issn = {0029-5582},
  doi = {10.1016/0029-5582(65)90862-X}
}

@article{MESHKOV1965199,
  title = {Validity of Many-Body Approximation Methods for a Solvable Model: ({{II}}). {{Linearization}} Procedures},
  author = {Meshkov, N. and Glick, A.J. and Lipkin, H.J.},
  year = 1965,
  journal = {Nuclear Physics},
  volume = {62},
  number = {2},
  pages = {199--210},
  issn = {0029-5582},
  doi = {10.1016/0029-5582(65)90863-1}
}

@article{chuStrongQuantumMetrological2023,
  title = {Strong {{Quantum Metrological Limit}} from {{Many-Body Physics}}},
  author = {Chu, Yaoming and Li, Xiangbei and Cai, Jianming},
  year = 2023,
  month = apr,
  journal = {Physical Review Letters},
  volume = {130},
  number = {17},
  pages = {170801},
  issn = {0031-9007, 1079-7114},
  doi = {10.1103/PhysRevLett.130.170801},
  urldate = {2026-07-16},
  langid = {english}
}

@misc{selsPrivateCommunication,
  author = {Sels, Dries},
  note   = {(private communication)}
}
\bibliographystyle{apsrev4-1}

\appendix


\prlsection{\bf Details of Universal Counterdiabatic Driving} The AGP is defined by the matrix elements in the instantaneous eigenbasis (setting $\hbar = 1$ henceforth):

\begin{equation} \label{eq:AGP_mtxelts}
    \langle m \vert A_\lambda \vert n\rangle = -i \frac{\langle m \vert \partial_\lambda H \vert n \rangle}{\omega_{mn}}
\end{equation}

\noindent where $\omega_{mn} = E_m - E_n$. 

Beyond the usual suspects of exactly solvable models or few-level systems , one generally cannot solve for (let alone implement experimentally) the operator $A_\lambda$ in Eq. \eqref{eq:AGP_mtxelts}. In many physical situations, one can find approximate solutions of Eq. \eqref{eq:AGP_mtxelts} variationally\cite{selsMinimizingIrreversibleLosses2017}. 

In particular, it has been observed that a natural ansatz for this variational method is an expansion in Krylov space~\cite{claeysFloquetengineeringCounterdiabaticProtocols2019,takahashiShortcutsAdiabaticityKrylov2024,bhattacharjeeLanczosApproachAdiabatic2023} via a series of nested commutators:

\begin{equation}
    \label{eq:krylov_AGP_comms}
    A_\lambda^{(\ell)} = i \sum_{k=1}^\ell \alpha_k \underbrace{[H,[H,..,[H,}_{2k-1} {\partial_\lambda H]]]}
\end{equation}

\noindent where $\ell$ defines the order of the expansion. As $\ell$ increases, we expect the performance of approximate CD driving to improve. This has matrix elements

\begin{equation} \label{eq:krylov_AGP_mtxelts}
    \langle m \vert A_\lambda^{(\ell)} \vert n\rangle = i \sum_{k=1}^\ell \alpha_k \omega_{mn}^{2k-1}\langle m \vert \partial_\lambda H \vert n \rangle
\end{equation}

\noindent where $\alpha_k$ are to-be-determined variational parameters. By comparing Eqs. \eqref{eq:AGP_mtxelts} and \eqref{eq:krylov_AGP_mtxelts}, one can see that trying to find an approximate counterdiabatic driving protocol $A_\lambda^{(\ell)} \approx A_\lambda$ is equivalent to trying to approximate $1/\omega_{mn}$ by odd polynomials in $\omega_{mn}$, with a weight function determined by the matrix elements $\langle m \vert \partial_\lambda H \vert n \rangle$. Within universal counterdiabatic driving~\cite{morawetzUniversalCounterdiabaticDriving2025a,finzgarCounterdiabaticDrivingPerformance2025a}, information about the system enters through the choice of a range of frequencies $[\omega_{min},\omega_{max}]$, such that the $\alpha_k$ are those which lead to a best fit of $1/\omega_{mn}$ uniformly within that window. In practice, there is an optimal ratio $\omega_{max} / \omega_{min}$ which is held fixed, so there is only one free parameter, $\omega_{max}$.

We highlight that that the approximation of $1/\omega_{mn}$ by odd polynomials can fail in two different regimes. First, when $\omega_{mn}$ is very small, since $1/\omega_{mn}$ diverges while the polynomials go to zero. This reflects the well-known difficulty in avoiding low-energy excitations when a system is driven. It can also fail when $\omega_{mn}$ is very large, so that $1/\omega_{mn}$ goes to zero but the polynomials diverge. Physically, this corresponds to the drive exciting additional high frequency transitions, and hence has been called the ``UV problem.'' These can be suppressed by either choosing $\lambda(t)$ to have the sufficiently rapidly decaying high-frequency tails in Fourier space \cite{finzgarCounterdiabaticDrivingPerformance2025a}, or by modifying a Floquet realization \cite{hatomura2026universaldigitizedcounterdiabaticdriving} of universal CD driving by developing an approximation scheme for the unitary evolution operator directly~\cite{selsPrivateCommunication}.


\prlsection{\bf Quantifying Sensitivity by Quantum Fisher Information} Here, we will briefly review the theoretical basis for quantum parameter estimation. In order to infer some underlying parameter of a system which is not directly observable, we need to define an estimator $\hat{E}_h$: an observable whose outcome can then be used to predict the underlying parameter $h$. If the system is in a $h$-dependent state $\vert \psi(h)\rangle$ and the estimator is unbiased, i.e. $\langle \hat{E}_h \rangle = h$, where $h$ is the true value of the parameter, one can rigorously show that the the variance of this estimator satisfies the quantum Cramèr-Rao bound: \cite{helstromQuantumDetectionEstimation1969, holevoProbabilisticStatisticalAspects2011} 

\begin{equation} \label{eq:QCRB}
    \langle \psi(h)\vert (\hat{E}_h - h)^2 \vert \psi(h)\rangle \geq \mathcal{F}_Q(h)^{-1}
\end{equation}

\noindent where $\mathcal{F}_Q$ is the Quantum Fisher Information (QFI) \cite{braunsteinStatisticalDistanceGeometry1994}. This is defined by

\begin{equation} \label{eq:QFI}
    F_Q(h) = 4 \left( \langle \partial_h \psi(h) \vert \partial_h \psi(h)\rangle - \vert \langle \psi(h) \vert \partial_h \psi(h)\rangle\vert^2 \right)
\end{equation}

Hence, a larger QFI means that a more precise estimate is possible. Connecting to the aforementioned Standard Quantum Limit, if the QFI grows faster than the number of particles $N$, then (at least in principle) there is some measurement which can be performed that yields sensitivity beyond the SQL. Growth of $F_Q(h) \sim N^2$ is known as Heisenberg scaling, and is the best achievable \cite{degenQuantumSensing2017}. We stress once again that the mere existence of such a measurement does not guarantee that it is straightforward to actually perform. Nonetheless, it is instructive to check what kind of sensitivity is possible, at least in principle.

\begin{figure*}
    \centering
    \includegraphics[width=\linewidth]{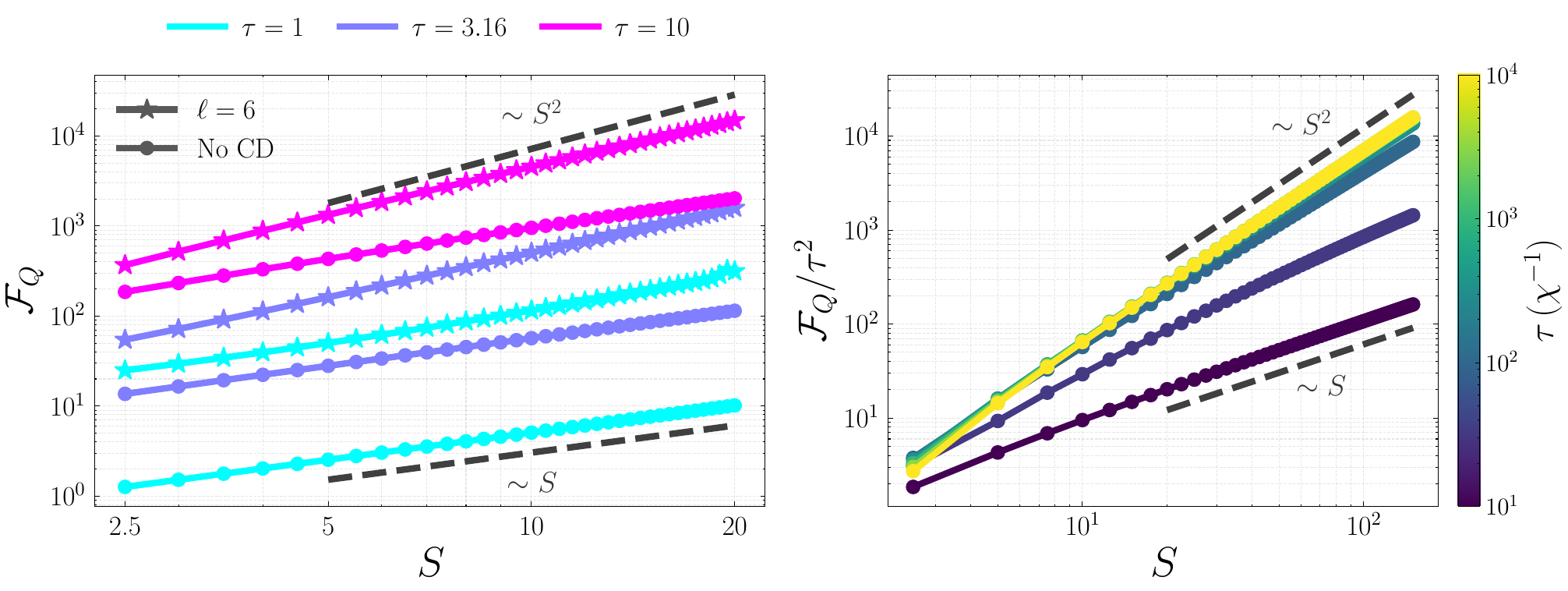}
    \caption{The Quantum Fisher Information, bounding the sensitivity of any possible measurement, for our sensing protocol. On the left, we see that for an appropriate choice of $\tau$ CD driving enables beyond-SQL sensitivity, agreeing with what we saw for a particular observable. For $\omega_{max} = 3.7$, this gives e.g. scaling $F_Q \sim S^{1.69}$ for $\tau \approx 4.0$. On the right, for long times where CD plays no role, we see that asymptotically as $\tau\to\infty$ the QFI scales quadratically in both $S$ and $\tau$. }
    \label{fig:QFI}
\end{figure*}

Quantitatively, we ask whether the Quantum Fisher Information exhibits beyond-SQL sensitivity -- i.e. if $F_Q(h) \sim S^\alpha$, with $1 < \alpha \leq 2$. To check this, we calculate $F_Q(h)$ numerically, defined in Eq. \eqref{eq:QFI}, where $\vert \psi(h)\rangle$ is the final state obtained after performing our preparation sensing protocol accelerated by universal CD driving. Derivatives are computed by finite differences.

The scaling of the QFI with $S$ is shown in Figure \ref{fig:QFI}. There are two features worth highlighting. Firstly, we can see that the presence of CD driving in our protocol leads to beyond-SQL sensitivity. This agrees with our results in Figure \ref{fig:SNR-results}. We also note that, in the limit of large $\tau$ where CD driving plays no role, the quadratic scaling of the QFI with both $\tau$ and $S$ is maximal \cite{giovannettiQuantumEnhancedMeasurementsBeating2004,boixoGeneralizedLimitsSingleParameter2007}. This scaling with $\tau$ is associated with maximum instability of long-time response to adiabatic deformations, which has been proposed as a classification of chaos \cite{Kim_2026,karve2026universaldynamicalresponseslow}.

\prlsection{\bf Performance of the Protocol in the Presence of Noise} As a simple demonstration of the performance of this sensing protocol in the presence of noise, we adopt the noise model used in Ref. [\onlinecite{norciaCavitymediatedCollectiveSpinexchange2018}]. There, the LMG model emerges effectively from many atoms interacting with a cavity photon mode which far detuned from resonance. The noise consists of a collective term (from cavity photon loss) and individual atomic dephasing, set by parameters $\Gamma$ and $\gamma_{el}$ respectively. The full Lindbladian dynamics is then given by

\begin{equation}
    \frac{d\rho}{dt} = -\frac{i}{\hbar}[H(\lambda),\rho]+ \mathcal{L}_\Gamma[\rho]+\mathcal{L}_{el}[\rho]
\end{equation}

\noindent where $H(\lambda)$ is defined as in the main text, but with $x \leftrightarrow z$ exchanged so as to be consistent with the quantization axis in Ref. [\onlinecite{norciaCavitymediatedCollectiveSpinexchange2018}]. The dissipative terms are given by

\begin{align*}
    \mathcal{L}_\Gamma[\rho]& =\frac{\Gamma}{2S}(2 S_- \rho S_+ - S_+ S_- \rho - \rho S_+ S_-) \\
    \mathcal{L}_{el}[\rho]& = \frac{\gamma_{el}}{2} (\sum_n \sigma_n^z \rho \sigma_n^z - \rho)
\end{align*}

\begin{figure*}
    \centering
    \includegraphics[width=\linewidth]{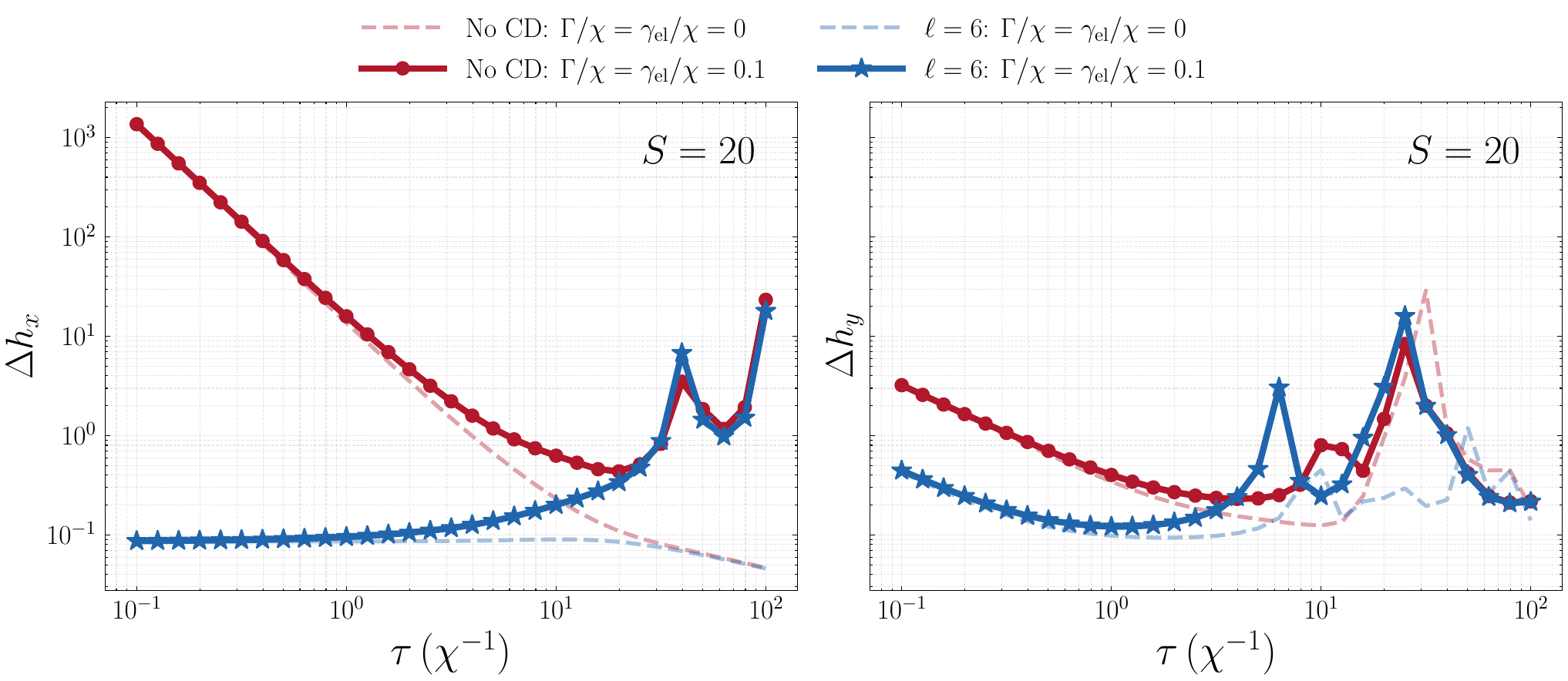}
    \caption{The fluctuations in the estimator of the field $h$ for different protocol durations, in the presence of weak noise. In the very slow, $\tau\to\infty$ limit, even weak noise destroys the sensitivity. On the other hand, if the noise is not too strong, there is still a strong signal in the regime $\tau \sim 1/\chi$ where the measurement of $\langle \hat{S}_y \rangle$ is most effective.}
    \label{fig:noisy_sim}
\end{figure*}

In Figure \ref{fig:noisy_sim}, we show the effect of the noise model on the sensitivity of our protocol. We see that even for relatively weak noise, trying to infer the field from a very slow protocol is totally infeasible. On the other hand, as long as the noise is not too strong, the effect of the field on the CD-accelerated state preparation protocol at $\tau \sim 1/\chi$ is not enough to destroy the sensitivity. We again highlight that a Floquet realization of such a CD protocol may provide even greater robustness to noise \cite{Boyers_2019} than in this direct commutator implementation.

\end{document}